\documentstyle[11pt,psfig,aas2pp4]{article}

\def\kms{\ifmmode \hbox{ \rm km s}^{-1} \else{ km s$^{-1} $}\fi} 

\def\ie{{\it i.e.}}
\def\ni{\noindent}

\slugcomment{to appear in AJ}

\lefthead{}
\righthead{The Spin of M87}

\begin{document}

\title{The Spin of M87 as measured from the Rotation of its Globular
Clusters}

\author{Markus Kissler-Patig \altaffilmark{1} and Karl Gebhardt \altaffilmark{2}
\affil{University of California Observatories / Lick Observatory, 
University of California, Santa Cruz, CA 95064}
\affil{Electronic mail: mkissler@ucolick.org, gebhardt@ucolick.org}
}

\altaffiltext{1}{Feodor Lynen Fellow of the Alexander von Humboldt Foundation}
\altaffiltext{2}{Hubble Fellow}

\begin{abstract}

We revisit the kinematical data for 204 globular clusters in the halo
of M87. Beyond 3 r$_{\rm eff}$ along the major axis of the galaxy
light, these globular clusters exhibit substantial rotation 
($\simeq 300\pm70$ km s$^{-1}$) that translates into an equally
substantial spin ($\lambda \simeq 0.18$). The present appearance of M87
is most likely the product of a single major merger, since this event is
best able to account for so sizable a spin. A rotation this large
makes improbable any significant accretion of material after this
merger, since that would have diluted the rotation signature.

We see weak evidence for a difference between the kinematics of the 
metal--poor and metal--rich population, in the sense that the metal--poor 
globular clusters appear to dominate the rotation. If, as we suspect, the last 
major merger event of M87 was mainly dissipationless and did
not trigger the formation of a large number of globular clusters, the
kinematic difference between the two could reflect their orbital
properties in the progenitor galaxies; these differences would be 
compatible with these
progenitors having formed in dissipational mergers. However, to put strong
kinematic constraints on the origin of the globular clusters themselves
is difficult, given the complex history of the galaxy and its last dominant 
merger event. 

\end{abstract}

\keywords{galaxies: individual (M87), galaxies: star clusters, galaxies:
elliptical and lenticular, cD} 

\section{Introduction}

Little is known about the kinematics of the halos of elliptical
galaxies. Unlike spiral galaxies, ellipticals contain little gas
which could serve as a kinematical tracer, and studies of the stellar light 
seldom extend further than 2 effective radii (r$_{\rm eff}$), 
due to the decreasing surface
brightness and the difficulty obtaining spectra of the diffuse stellar light. 
Furthermore, companion galaxies, which could serve as kinematical probes
in the outer halo, are usually not numerous enough and too distant 
to probe the range between 2 and 10 (or more) r$_{\rm eff}$.
Recently, however, globular clusters and planetary nebulae have become popular test 
particles to probe the kinematics in the halo of giant ellipticals.
With the commissioning of 8m--class telescopes, large
samples of radial velocities can be obtained, extending out to several r$_{\rm
eff}$ from the center of the galaxies. Globular clusters are especially
numerous around central giant ellipticals and therefore are probably the
best tracers of the outer kinematics in these galaxies. 

The kinematics derived from the globular clusters can be used to
constrain galaxy formation history. For example, cosmological
N--body simulations make specific predictions of the amount of spin that
will result from the torques caused by the multiple accretion of objects
(see Sect.~4.2); moreover, a formation via dissipational mergers
predicts both varying kinematics of the globular clusters with radius
and different kinematics between the newly formed globular clusters and
the ones brought in by the progenitors (see Sect.~4.3).

Thus, the kinematics of a large sample of globular clusters provides not only 
constraints on the formation and evolution of the galaxy, but also
a better understanding of the formation of the globular cluster
systems. In the central cD galaxy in Fornax (NGC 1399), the kinematics
of the globular clusters constrain the origin of the large 
over--abundance of globular clusters (Kissler-Patig et al.~1998a). In
the other dominant giant elliptical in Virgo (NGC 4472), the kinematics
of the globular clusters suggest a formation via two massive gas--rich
galaxies (Sharples et al.~1998). 

Here, we revisit the central giant elliptical in the Virgo galaxy
cluster, M87 (NGC 4486), and derive the kinematics in a region between 1 and 5
r$_{\rm eff}$ using published radial velocities for over 200 globular
clusters (Cohen \& Ryzhov 1997). Previous work on the kinematics of
globular clusters around M87 also include studies from Huchra \&
Brodie (1987), Mould et al.~(1990), and Brodie \& Huchra
(1991). These studies were based on 20 to 45 objects.
Interestingly, Mould et al.~reported dynamically significant rotation in
both M87 and M49. In M87 their study extends to $\simeq 400\arcsec$
radius; they report a rotation along the major axis of $0.60\pm0.27$
km s$^{-1}$ arcsec$^{-1}$, i.e., $240\pm108$ km s$^{-1}$ at their extreme radius.
The extension of this result is the main motivation for this work.

In the next section (Sect.~2) we briefly describe 
the sample and our new analysis before presenting in Sect.~3 the velocity 
dispersion and rotation velocity both as a function of radius and an
estimate for the dimensionless spin parameter $\lambda$ for the galaxy. 
We discuss our findings and present our conclusions on the formation
history of M87 in Sect.~4. 

\section{Globular cluster sample and data analysis}

\subsection{The sample}

We used the sample of 230 globular cluster candidates around M87 for which 
Cohen \& Ryzhov (1997) published radial velocities, including an
additional cluster (ID 682) and the update on five velocities given in Cohen et
al.~(1998). The coordinates were obtained from Strom et al.~(1981). 
Following Cohen \& Ryzhov, we excluded galactic
stars by selecting a sub--sample of 204 globular clusters with
velocities $v_r > 250$ km s$^{-1}$. We also excluded an additional candidate at
$v_r=350$ km s$^{-1}$. The velocity offsets from the systemic velocity
for the remaining objects are well below the escape velocity of the galaxy. 
Also, we do not expect any torques or tidal
influence on the velocities from the Virgo cluster, since M87 is located at its
exact center (e.g., Schreier et al.~1982).  The resulting sample is shown in
Fig.~1, where we plot the position of all of the globular clusters.
The symbol sizes reflect the difference between the cluster velocities
and our adopted systemic heliocentric velocity for M87 of $1282\pm9$ km
s$^{-1}$, taken from the RC3 (de Vaucouleurs et al.~1991). Open circles show
approaching and filled circles show receding globular clusters; the dashed
line marks 1 $r_{\rm eff}$ ($\sim 100\arcsec\ $, e.g., Goudfrooij et al.~1994) 
of the galaxy light and the solid line marks 3 $r_{\rm eff}$.

\subsection{Analysis}

Cohen \& Ryzhov (1997) comprehensively present and discuss the
velocity dispersion of the globular clusters as a function of distance
from M87. The authors also discuss rotation, but only divide their
data into two radial bins (within and outside 180\arcsec\ ), and find no
significant deviation in these two sub--samples compared with the
rotation of the whole sample: about 100 km s$^{-1}$ along a position
angle (measured from north through east throughout this paper) 
of $\sim 150^{\circ}$, defined as the angle of maximum positive rotation.

We extend their study by investigating the radial profiles of the
rotation, position angle, and velocity
dispersion. For each globular cluster, we use the nearest 75 data
points in radius to measure the above
three parameters, allowing this sub--sample to drop to 30 data points
at the extreme radii, 80\arcsec\ and 400\arcsec . 
For each sub--sample, a maximum--likelihood fit
determines the position angle, rotation amplitude, and dispersion at that
radius (see Pryor \& Meylan 1993, for example, for the basic approach). 
The fit measures the best sinusoid for the velocity data as a function
of position angle. We fix the mean velocity of the
sub--samples to be equal to the global mean velocity. The amplitude of
the fit provides the rotation, the angle of maximum positive rotation
provides the PA, and the standard deviation about the sinusoid
provides the velocity dispersion. The uncertainty in the fit is
derived in two ways: using the classical covariance matrix and using a
bootstrap.   For the bootstrap
uncertainties, at each location where a cluster exists, we draw a point
from a Gaussian distribution with the mean given by the 
rotation amplitude and PA at that radius, and
the standard deviation given by the dispersion at that radius. A
velocity uncertainty of 50 km s$^{-1}$ is also included. This sampling
provides one realization and we similarly fit the three
parameters. Generating 100 realizations then allows for a distribution
of values at every radius; using the 16th and 84th percentile values
provide the 68\% confidence band. The 68\% confidence band and the
classical $1\sigma$ uncertainties are similar at each radius; in all
of the figures we plot the confidence band based on the bootstrap
technique.

As mentioned above, each realization results from the assumption of a
gaussian distribution in the velocity differences from the rotational
velocity. We have also used the actual differences for the boostrap
simulations; i.e., we randomly draw with replacement from the
distribution of differences to generate a simulated dataset. There is no
difference in the results using either a Gauss or the actual distribution.

Cohen \& Ryzhov (1997) quote a error of 100 \kms\ for their
velocities, but suggest that this is an overestimation; we
therefore use an uncertainty of 50 \kms. However, we verirified that
there is no difference in the results for either uncertainty, 
as is expected since the high velocity dispersion (about 370 \kms) dominates the
uncertainty in the estimation of the rotational velocity. 

Our results are shown in Fig.~2 and 3, where the projected rotation
$v\cdot sin(i)$, projected velocity dispersion $\sigma_v$, position
angle, and $v\cdot sin(i) / \sigma_v$ 
(with their associated 68\% confidence bands) are plotted against radius.
Figure 2 shows the results when the PA is allowed to vary and
Fig.~3 shows the results for a fixed PA at 120$^\circ$ (the mean value
for the PA past 3 r$_{\rm eff}$). Figure 4 plots the individual velocity 
measurements versus position angle in four radial bins of equal width
(100\arcsec) with mean radii of 1, 2, 3, and 4 r$_{\rm eff}$, together with
the derived rotation at these radii; rotation is clearly present in
the outermost radial plot. Using Monte--Carlo simulations and given a
velocity dispersion of 370 \kms, the probability of measuring a rotation 
this large, when no rotation is present, is less than 0.1\%.

\section{Kinematics of M87 between 1 and 5 r$_{\rm eff}$}

Within 1 r$_{\rm eff}$, there is only marginal evidence for rotation
in the sample, a result compatible with the lack of any significant rotation of
the stars (Javis \& Peletier 1991, Sembach \& Tonry 1996 and
references therein).  At $\sim 1.5$ r$_{\rm eff}$, a group of globular
clusters rotating along an axis offset by $\simeq 60^{\circ}$ from the
major axis possibly exists; however, the result is only marginally
significant.  We will come back to this point in Sect.~3.2.  The main
feature is the monotonic increase of rotation past 2 r$_{\rm eff}$.

\subsection{Kinematics beyond 2 r$_{\rm eff}$}

Beyond 2 r$_{\rm eff}$, the globular clusters rotate roughly along the
major axis of the galaxy ($\simeq 150 ^{\circ}$, e.g., Goudfrooij et
al.~1994), with higher velocities in the SE. The position angle remains
stable and the rotation amplitude increases from $\simeq 50$~\kms\ at
2 r$_{\rm eff}$ to $\simeq 300$~\kms\ at 4 r$_{\rm eff}$.  The
velocity dispersion remains constant at $\simeq 370$~\kms\ from 2 to 4
r$_{\rm eff}$. The increase in the velocity dispersion at the largest
radii reported by Cohen \& Ryzhov (1997) is likely due to increasing
rotational velocity.  Accordingly, $v\cdot sin(i)/\sigma_v$ increases
with radius from values around 0.2 to a value close to 0.8 at 4
r$_{\rm eff}$.  Further evidence in support of the rotation is the
correspondence of the major-axis position angle of the galaxy
isophotes to the position angle of the maximum rotation. In addition,
the amount of isophotal flattening is consistent with an isotropic
oblate rotator given the amplitude of $v/\sigma$. For an oblate rotator, we
would expect an ellipticity of $\epsilon \simeq 0.25$ for the $v /
\sigma_v \simeq 0.5$ seen at $\simeq 350\arcsec $, and
$\epsilon \simeq 0.45$ for the $v / \sigma_v \simeq 0.9$ seen at
$\simeq 400\arcsec\ $ (Binney \& Tremaine 1987).

McLaughlin et al.~(1994) studied in detail the spatial structure of
the globular clusters in M87 and found the system to be elliptical,
aligned along the major axis, with the ellipticity $\epsilon$
increasing steadily to values of $\simeq 0.3$ at about 300\arcsec. The
diffuse stellar light was studied to even larger radii (Liller 1960,
King 1978, Carter \& Dixon 1978).  These studies posit similar
findings; the stellar light of M87 is flattened 
with an ellipticity smoothly increasing to values of $\epsilon$
between 0.35 and 0.4 at 350\arcsec\ -- 400\arcsec. These observations
strongly support the rotation and its amplitude detected in the
globular clusters at these large radii.

In order to compare our rotation to various theoretical predictions in
Sect.~4, we calculate the dimensionless spin parameter (Peebles 1971)
$\lambda=J\ E^{1/2}\ G^{-1}\ M^{-5/2}$, where $J,E$ and $M$ are the
angular momentum, total energy, and total mass of the galaxy
respectively, and $G$ is the gravitational constant. $\lambda$
measures the total spin of a galaxy; however, since globular cluster
radial velocities only exist over a limited radial range, we only measure a
fraction of the total spin. We must therefore either extrapolate to
include the whole galaxy or use the $\lambda^\prime$ parameter
introduced by Barnes (1992). $\lambda^\prime$ is given by $J/J_{\rm
max}$, where $J$ is the angular momentum of the subsample and $J_{\rm
max}$ is the maximum angular momentum that the subsample can
have. These two spin parameters are not directly comparable; \ie, a cold
disk has $\lambda=0.43$ and $\lambda^\prime=1.0$ (Barnes 1992). Below
we calculate both parameters for comparison with theoretical predictions.

To estimate the spin parameters, we deproject the velocity profile
with various assumptions, use the logarithmic potential given by Weil
et al.~(1998), and thus compute the mass, binding energy, and angular
momentum as functions of radius.  For the logarithmic potential we use
$v_0=550$ km s$^{-1}$, $R_c=100\arcsec $, and $q=0.85$, corresponding
to a total mass of M87 of $M(R=120 {\rm kpc})\simeq 1.2 \times 10^{13}
M_\odot$. For the deprojection of the rotational velocity we use the
Abel equation, and the surface brightness and density profile
of Weil et al.  Assuming that the galaxy is edge--on, we
numerically integrate

\begin{equation}
\nu(r)v_{\phi}(r) = -{r\over\pi} \int_r^\infty {{d\over dx}\left({1\over x}
\Sigma(x)v_p(x)\right)} {dx\over\sqrt{x^2-r^2}},
\end{equation}

\noindent to obtain the rotational velocity along the major axis. Where $\nu$ is
the density used in Weil et al., $v_{\phi}$ is the internal rotational
velocity, $\Sigma$ is the surface brightness profile (given in
Weil et al.), and $v_p$ is the projected rotation in Fig~3. The
integral extends to $r=\infty$, but our data only goes out to
400\arcsec , thus requiring an extrapolation.
We use four different extrapolations beyond our last
data point: zero rotation outside of 400\arcsec , rotation decreasing
linearly with increasing radius (by 0.375 km s$^{-1}$ arcsec $^{-1}$), 
constant rotation, and rotation increasing linearly with increasing
radius (by 0.625 km s$^{-1}$ arcsec $^{-1}$).

Two more assumptions are necessary to obtain the total angular momentum: 
the relation of the globular cluster's rotation profile to the
galaxy rotation profile and the 2--D structure of the velocity field.
We assume that the galaxy has the same rotation profile as the globular 
cluster system. 
For the velocity field, we assume constant rotation on cylinders and 
obtain the rotation at every position in the galaxy from the major axis
rotation profile. The total angular momentum for the galaxy is then given by

\begin{equation}
J(r) = 2\pi \int_0^r r^{\prime 2} dr^\prime \int_0^{\pi} v_{\phi}(R)~
R~ \rho(R,z)~ sin\theta d\theta,
\end{equation}

\ni where $R=r^\prime sin\theta$ and $z=r^\prime cos\theta$. 
Similarly, we substitute $v_{\rm max}(R)$
for $v_{\phi}(R)$ in equation~2 to obtain the maximum angular
momentum, $J_{\rm max}(r)$. For $v_{\rm max}(R)$, we use
$\sqrt{v_{\phi}^2(R)+\sigma^2}$, where $\sigma=370$\kms, the measured
projected velocity dispersion. The calculated $v_{\rm max}$ is very
similar to the circular velocity at all radii, and using either one
provides similar values for $J_{\rm max}$.  $J/J_{\rm max}$ equals
$\lambda^\prime$.

Figure 5 plots $\lambda^\prime$ as a function of radius for the
various assumptions. The confidence bands are calculated in the same
manner as in Figure~4. In the extreme case, where the rotation
velocity drops to 0 beyond our last observed point, $\lambda^\prime$
rises steeply to values around 0.8 past 300\arcsec . In the more
realistic cases, where we assume that the points at the largest radii
are near the maximum net rotation and that the rotation curve remains
approximately flat beyond our last point, $\lambda^\prime$ reaches
values of $> 0.10\pm0.05$. To decrease $\lambda^\prime$ even
further at $r<400$\arcsec , we would have to assume that the rotation 
velocity continues
to rise steeply beyond 400\arcsec\ to values well above 700~\kms, an
unrealistic probability. For any assumption, however, $\lambda^\prime$ obtains
a high value; putting all of the rotation at $r<400$\arcsec\ results
in $\lambda^\prime$ around 0.8, whereas allowing the largest rotation
to be at $r>400$\arcsec\ suggests a lower $\lambda^\prime$ for the
inner radii, but $\lambda^\prime \simeq 0.7$ at larger radii. The
significant rotation seen around $r=400$\arcsec\ ($v/\sigma = 0.8$)
forces M87 to obtain high values of $\lambda^\prime$ {\it independent} of
the rotational--velocity extrapolation used.


Equation 2 is used to calculate $\lambda^\prime$ integrated from zero to $R$.
However, $\lambda^\prime$ can also be computed in radial bins
(e.g.~Hernquist \& Bolte 1992). Figure~6 plots the 
local $\lambda^\prime$ integrated over a
40\arcsec\ region around $R$. 

Note that we make several important assumptions. First, we speculate
that M87 is seen edge--on, which could lead to an under--estimation of
the derived angular momentum (by
$\simeq 15$\% for $i=60^\circ$).  However, the amount of flattening
in the isophotes for M87 at large radii ($\epsilon \simeq 0.4$)
implies that it is nearly edge--on, since any inclined orientation
would require an unlikely more flattened system. Our second assumption
is that the rotational velocity of the globular clusters represents
the rotation velocity of the total mass.  If the globular clusters
rotate twice as fast as the rest of the mass --- an extreme case when
compared to results for rotation velocities of galaxies in
cosmological N--body simulations, see Sect.~4 --- we would overestimate
$J$, i.e.~$\lambda^\prime$, by a factor of two. Simulations, however,
tend to show that the globular cluster system rotates more slowly than the 
stellar body (Hernquist \& Bolte 1994). Finally, we use a
mass normalization for the logarithmic potential of Weil et al.~(1998)
of $v_0=550$~\kms; using extreme values as low as $v_0=400$~\kms\ or as
high as $v_0=650$~\kms\ results in values of $\lambda^\prime \simeq
0.2$ and $\simeq 0.05$ respectively at 400\arcsec .

As well as $\lambda^\prime$, we also estimate $\lambda$. For
$\lambda$, we need the mass, angular momentum, and the binding
energy. We cannot use $\lambda$ as measured from a subset of a bound
system since $\lambda$ measures the total spin for a bound system; we
must therefore extrapolate to large radii to compare our measured
$\lambda$ to in the theoretical predictions discussed
below. The theoretical calculations generally use objects out to 1--2
half-mass radii to estimate the spin parameter. For M87, the half mass
radii is anywhere from 150--300~kpc (assuming a distance of 16.5 Mpc), 
depending on where one truncates
its halo. The globular cluster velocity data only extends to around
35~kpc, so we must make an enormous extrapolation to determine the
angular momentum. For the mass and binding energy, there is no
extrapolation since the mass model incorporates the data at these
radii. The integral over the density times the potential yields the
binding energy, and the integral of the density provides the mass
profile. We extrapolate the net rotation by assuming that it is constant
between our last point and the half-mass radius, and Equation~2 then 
provides the measurement of the angular momentum. With this extrapolation,
$\lambda$ asymptotes to a value of 0.18 at a radius around 600\arcsec\
(60~kpc) and stays constant beyond there.

In summary, our best values for the dimensionless spin parameters of
M87 are $\lambda^\prime \simeq 0.7$ at $r>600$\arcsec\ and
$\lambda \simeq 0.18$ for the whole system. We will discuss the implications
for the formation of M87 in Sect.~4.

\subsection{The kinematics around 1.5 r$_{\rm eff}$}

Two groups of clusters with 4--5 members each --- immediately outside 1
r$_{\rm eff}$ to the North and South of the galaxy center (see Fig.~1)
--- appear to produce the rotation seen at $\simeq 1.5$ r$_{\rm eff}$
around an axis offset from the rotation axis at large radii.  This
pattern does not extend further out. The statistical significance is a
little greater than 1 $\sigma$ and still compatible within the errors with
a constant low rotation out to 2
r$_{\rm eff}$ along the major axis. The offset axis may, however, be
due to incomplete azimuthal coverage of the data and, in particular,
the lack of data in the NW around 2 r$_{\rm eff}$.  Alternatively,
this rotation could be due to a separate group of clusters on orbit
around M87 (e.g., from an accreted system).  Abundances are available
for 8 of these clusters (Cohen et al.~1998). The values scatter over 
75\% of the range spanned by the full sample (${\rm Mg}_{\rm b}<3$\AA), 
neither supporting nor excluding the accretion of a dwarf companion. The exact 
nature of these clusters remain to be investigated, but do not influence our
results beyond 2 r$_{\rm eff}$.

\section{Constraints on the formation of M87}

We compare the kinematical properties of the globular clusters to
galaxy formation scenarios, and briefly discuss why our results can
not be explained by a single infalling satellite.  From hierarchical
formation scenarios the high value derived for the spin parameter suggests 
that the evolution of M87 was likely dominated by {\it one} major merger event.

\subsection{Only an infalling satellite?}

We must verify that the measured rotation is characteristic of the
system and not a group of globular clusters associated with a
satellite or originating from a tidal tail of an interacting, stripped
galaxy. Evidence against these latter hypotheses is the  
alignment of the globular clusters with the major axis, the match with
the stellar rotation, and the relatively smoothly increasing rotation
and ellipticity of the whole system. Weil et al.~(1998) recently
reported diffuse (28 mag arcsec$^{-2}$) stellar light around M87 and
suggest that it results from the accretion of a small spheroidal
galaxy. The fan of light nearly aligns with the major axis with a
large opening angle, suggesting that the orbit must have passed close
to the center. However, their simulations show that the accreted
galaxy must have been of low mass ($10^9 - 10^{10} M_\odot$), and the
associated number of globular clusters would be small.  By comparison
with numbers of globular clusters per unit mass for similar galaxies in the 
compilation of Zepf \& Ashman
(1993), the total number of globular clusters associated with such a
galaxy would be of the order 3 to 30; the number present in our
magnitude limited sample would be a factor of 3 to 4 lower. Thus, it is
not possible for these accreted clusters to dominate the globular
cluster system of M87 at 4$_{\rm eff}$.

Three other dwarf galaxies are potential candidates for satellites:
NGC\ 4486A, NGC\ 4486B, and IC\ 3443.  They are projected within
10\arcmin\ of the center of M87 and within $30^{\circ}$ of the PA of the
major axis and could be at a similar distance as the central giant
elliptical.  Both NGC\ 4486A and NGC\ 4486B are counter--rotating with
respect to the stars and globular clusters with velocities differences 
of $\simeq 800$ and $\simeq 200$ km s$^{-1}$, respectively, compared to the
systemic velocity of M87.  IC\ 3443 is a dwarf galaxy with 
$M_{B_T}\simeq -15.5$ and it seems unlikely that it could have
contributed many globular clusters to M87.  Finally, we checked the
abundances published by Cohen et al.~(1998) for the clusters which
dominate the rotation at large radii. No systematic pattern is
visible; the values scatter over the range spanned by 75\% of the range 
of the full sample, on the metal--poor side.

In summary, no companion can significantly contribute to the globular
cluster system of M87 at $\simeq$4$_{\rm eff}$, so the rotation must
be intrinsic to the globular cluster system associated with M87.

\subsection{An explanation for the high spin parameter}

Peebles (1969, 1971) first estimated the resulting angular momentum of
a galaxy formed by gravitational instability. More recently
(e.g.~Ueda et al.~1994 and references therein), specific predictions
were made for hierarchical clustering models with large cosmological
N--body simulations.  The results of the different simulations agree
well; all simulations appear to predict a lower dimensionless spin
parameter $\lambda$ than we observe in M87. Values for $\lambda$ vary
between 0.01 and 0.1 with a mean around $\lambda=0.05$ and are mostly
insensitive to cosmological parameters and fluctuation spectrum
shape. In addition, $\lambda$ tends to be lower in high--density
environments and appears to be anti--correlated with galaxy mass (Ueda
et al.~1994, Efstathiou \& Jones 1979, Barnes \& Efstathiou 1987).

The spin parameter that we derive for M87 is $\lambda \simeq
0.18$. 
The result is somewhat sensitive to our different assumptions, but
should not be off by more than a factor of two (see Sect.~3.1).
That is, the result for M87 is only
marginally consistent with the simulations, and lies at the upper end
of, or above, the simulation results.

Interestingly, Warren et al.~(1992) noticed in their simulations the
apparent decrease of $\lambda$ at masses comparable to M87, which they
attributed to their finite computational volume rather than to a
physical process: the most massive halos in simulations have no
similar large halos to tidally torque them. That is, M87 would have
to have encountered a equally massive galaxy in order to gain enough
angular momentum to explain values of $\lambda$ around 0.1 or
higher. High values of $\lambda$ have also been borne out in other types of 
simulations.  Hernquist (1992, 1993) showed in merger simulations of
two equal--sized galaxies how the spin parameter could increase in the
outer halo. Hernquist \& Bolte (1992) made specific predictions for
globular clusters at different radii in the end--product of such a
merger and noticed that $\lambda^\prime$ rises from values around 0.05 in the
center to values as high as 0.2--0.3 at several effective radii.

These simulations seem to indicate that the most likely scenario is
that M87 gained its large amount of angular momentum through a merger
of two large galaxies of about equal mass. This scenario does not
exclude a hierarchical formation for the progenitor galaxies or further
accretion of smaller companions (see Sect.~4.1), but {\it the present
appearance of M87 must have been created by one single major merger
event}.  It seems very unlikely that several merging galaxies
would fall into M87 on the same orbital path so as to not dilute
each other's rotation signatures.  Given the total stellar 
mass of M87 (several $10^{12}
M_{\odot}$), the two progenitors must have had stellar masses around $10^{12}
M_{\odot}$ or above, which effectively rules out that M87 is the
product of the merger of two (several $10^{11} M_{\odot}$) spiral
galaxies.  The most likely progenitors are two giant ellipticals,
perhaps central dominant galaxies in their own groups, given the large
masses involved. In agreement with this, M87 shows no significant recent burst
of star formation; the last major merger appears to have been
gas--poor or mainly dissipationless, unless it happened well before
$z=1$.

\subsection{A comparison with NGC 4472 and NGC 1399}

We carried out a similar analysis for two other giant elliptical
galaxies: NGC 4472 (M49), the brightest galaxy in the Virgo galaxy
cluster, for which we used 57 globular cluster radial velocities
compiled by Sharples et al.~(1998); and NGC 1399, the central cD
galaxy in the Fornax galaxy cluster, for which we used 74 globular
cluster radial velocities compiled by Kissler-Patig et al.~(1998). We
only summarize the results here since we feel that the samples are
too small to draw strong conclusions. However, we note that in both
cases the rotation at large radii is significantly smaller than in M87,
resulting in smaller spins. 

In NGC 4472 the clusters out to $\simeq 1.5 r_{\rm eff}$ have
significant rotation, driving the value of $\lambda^\prime$ up to $\simeq
0.5$, but it decreases to $\simeq 0.1$ at the last data point
(around 2.5 $r_{\rm eff}$) and, extrapolating with constant rotation
amplitude, $\lambda^\prime$ remains approximately constant out to larger 
radii.
The total spin parameter $\lambda$ around the virial radius is low,
around 0.01. The result for $\lambda^\prime$ is very sensitive to a particular 
group of $\simeq 10$ objects at small radii. 

In NGC 1399 no significant
rotation could be detected in the inner regions and only small rotation 
($\simeq 150\pm 100$ km s$^{-1}$) was measured past 300\arcsec\ ($\simeq
6 r_{\rm eff}$). Accordingly, we derive values of $\lambda^\prime$
around 0.3, but compatible with 0. The total spin parameter
$\lambda$ near the virial radius lies around 0.05. 
These derived values for the spin parameters are in good agreement with
the values obtained by cosmological N--body simulations for the formation of 
central cD galaxies.

\section{Differences between the red and blue globular clusters}

\subsection{Kinematics of the globular cluster sub--populations}

The globular cluster system of M87 is known from photometry to host
two distinct population of globular clusters (Whitmore et al.~1995,
Elson \& Santiago 1996). In order to investigate whether these two
sub--populations differ in their kinematics, we divide the kinematic
sample according to the metallicity obtained spectroscopically for 150
candidates by Cohen et al.~(1998). The ``blue'' and ``red''
populations are defined as globular clusters with [Fe/H]$<-0.9$ (80
candidates) and [Fe/H]$>-0.9$ (70 candidates), respectively. This cut is 
motivated
by the gap at similar metallicities in the globular cluster population
of the Galaxy (e.g.~Zinn 1985) and the metallicity corresponding to
the gap in the $V-I$ color distribution of the globular clusters in
M87 (e.g.~Whitmore et al.~1995, Elson \& Santiago 1996).  The analysis
was performed in the same manner as for the full sample (see
Sect.~2). Figure 7 shows the results using a fixed position angle
($120^{\circ}$): velocity dispersion $\sigma_v$ and rotation velocity
$v\cdot sin(i)$ of the blue (thick solid line) and red (thin solid
line) samples are plotted as a function of radius. The 68\% confidence
bands (dotted lines) are similar for both samples and are shown only
for the blue sample.

Blue and red samples have similar mean velocities ($1299\pm44$ km
s$^{-1}$ and $1292\pm47$ km s$^{-1}$ respectively).  The rotation in
the red sample seems constant (and compatible with no rotation) at all
radii, while the rotation of the blue cluster population seems to
increase with radius past 200\arcsec\ and is significant beyond
300\arcsec . However, given the sample sizes, the rotation amplitudes of
the two samples only differ by $\simeq 1 \sigma$.

The blue and red samples have similar velocity dispersions at all
radii, however, their rotation profiles differ, implying different
total $V^2 \equiv (\sigma ^2 + v_{\rm rot}^2)$ profiles. Since both groups
trace the same mass distribution, either their radial density
profiles are different (e.g., one population has a flatter density
profile than the other) or their families of orbits are very different
(e.g., the globular clusters of one population having either primarily
tangentially or radially biased orbits). Strom et al.~(1981) first
noticed that the mean color of the globular cluster system in M87
becomes bluer with increasing distance from the center. Neilsen et
al.~(1998) suggest that the color variation is due to a changing
number ratio of blue to red clusters with radius (similar to what was
seen in NGC 4472 by Geisler et al.~1996, and NGC 1380 by Kissler-Patig
et al.~1997), suggesting different density profiles.

At 400\arcsec , the ratio of the total $V^2$ of the two samples is
$\simeq 1.4$. Assuming isotropic orbital distributions, in order for
the two populations to trace the same potential, this ratio translates
into a ratio of $\simeq 1.4$ for the exponents of the density
profiles. This ratio does seem typical for the exponents of the
density profiles of the blue and red populations (e.g., the cases of
NGC 4472 and NGC 1380) and for the ratio of the exponents for those globular
cluster populations which follow the stellar light profiles and
those whose density profile is more extended than the stellar light
(see the compilation in Kissler-Patig 1997, for example).

Similar velocity dispersion profiles for the red and blue populations
are therefore compatible with the larger rotation of the blue globular
clusters if the red population has a steeper density profile, which
also explains the color gradient in the system.

\subsection{Constraints on the formation of the globular clusters}

M87 is known to host an extremely high number of globular clusters with
respect to its mass (e.g., McLaughlin et al.~1994 and references
therein), as well as hosting at least two different globular cluster
sub--populations.

Several mechanisms can explain the presence of different globular
clusters populations (e.g., Kissler-Patig et al.~1998b). In the case of
M87, the last merger probably did not involve a large amount of star
and globular cluster formation; a result supported by the findings of
Cohen et al.~(1998), who showed that the vast majority of globular
clusters are old and must have formed in earlier events. In this case,
the globular cluster kinematics cannot strongly discriminate between
different formation scenarios since they were dominated by this last
merger.  Indeed, there is only weak evidence that the blue globular
clusters cause most of the net rotation. We also stress that the last major
merger cannot explain the globular cluster over--abundance and point
to Harris et al.~(1998) and C\^ot\'e et al.~(1998) for alternative
scenarios in the case of M87.

It is possible that the apparent concentration of the net rotation in 
the blue globular clusters may be {\it a relic from the situation in the
progenitor galaxies}.  Hernquist (1993 and references therein) showed that, in
an equal--mass merger, the angular momentum of rotating components
tends to be conserved.  In the case of M87, the observations
would be compatible with formation of the progenitors by
dissipational mergers (see Ashman \& Zepf 1992). The red globular clusters
would have formed from the infalling gas that accumulated in the
center of the progenitors and would have little angular momentum
(e.g., Barnes \& Hernquist 1996). Whereas the older blue globular
clusters would have been partly spun up by the mergers which formed
the progenitors of M87.  But the orbital mixing caused by the last major
merger event in M87 makes it difficult to draw firm conclusions on the
orbital history of its globular clusters from their present--day
kinematics.

\section{Summary}

We have re--analyzed published radial velocities for 204 globular clusters around
M87 and found significant rotation in the outer regions ($>
250\arcsec $).  We derive a dimensionless spin parameter for M87 of
$\lambda \simeq 0.18$ ($\lambda^\prime \simeq 0.7$) 
from the rotation of the globular
clusters.  A comparison with cosmological N--body simulations, argues
that such a high spin parameter is most likely the product of an equal--mass
merger. A single major merger must have been the dominant event shaping
the kinematics and appearance of the galaxy.  There is
some evidence that the rotation is confined to the metal--poor
globular clusters. If, as assumed, the last merger was mainly
dissipationless, this kinematic difference could reflect the situation
in the progenitor galaxies of M87.

\acknowledgments

We thank Dean McLaughlin, Lars Hernquist, Joel Primack, and Tsafrir Kolatt 
for interesting discussions, and are thankful to J.G.~Cohen,
A.~Ryzhov and J.P.~Blakeslee for having made available electronic
versions of their results. We are further grateful to Carlton Pryor 
for a careful reading of the manuscript.
MKP is supported by a Feodor Lynen
Fellowship of the Alexander von Humboldt Foundation. KG is supported
by NASA through Hubble Fellowship grant HF-01090.01-97A awarded by the
Space Telescope Science Institute, which is operated by the
Association of the Universities for Research in Astronomy, Inc., for
NASA under contract NAS 5-26555.

\clearpage
 
\begin{deluxetable}{r c c r}
\footnotesize
\tablecaption{Rotation velocity and velocity dispersion as a function of
radius}
\tablewidth{0pt}
\tablehead{
\colhead{Radius} & \colhead{Rotation velocity} 
& \colhead{Velocity dispersion} & \colhead{Spin} \nl
\colhead{arcsec} & \colhead{$v\cdot sin(i)$ [km s$^{-1}$]} & 
\colhead{$\sigma_v$ [km s$^{-1}$]} &  \colhead{$\lambda^\prime$} 
} 
\startdata
 83 & $ 56 _{- 38} ^{+ 43}$ & $333 _{- 37} ^{+ 34}$ & $0.06 _{-0.04}
 ^{+0.06}$ \nl
 93 & $ 13 _{- 13} ^{+ 43}$ & $318 _{- 29} ^{+ 37}$ & $0.04 _{-0.03}
 ^{+0.05}$ \nl
103 & $ 18 _{- 18} ^{+ 44}$ & $319 _{- 28} ^{+ 38}$ & $0.04 _{-0.03}
^{+0.06}$ \nl
114 & $ 77 _{- 37} ^{+ 39}$ & $326 _{- 31} ^{+ 31}$ & $0.04 _{-0.03}
^{+0.07}$ \nl
123 & $ 42 _{- 35} ^{+ 41}$ & $336 _{- 28} ^{+ 29}$ & $0.06 _{-0.04}
^{+0.07}$ \nl
134 & $ 81 _{- 37} ^{+ 43}$ & $350 _{- 29} ^{+ 30}$ & $0.08 _{-0.05}
^{+0.06}$ \nl
143 & $ 71 _{- 45} ^{+ 53}$ & $349 _{- 28} ^{+ 32}$ & $0.09 _{-0.06}
^{+0.07}$ \nl
154 & $ 56 _{- 45} ^{+ 45}$ & $335 _{- 26} ^{+ 30}$ & $0.12 _{-0.06}
^{+0.07}$ \nl
164 & $ 62 _{- 47} ^{+ 49}$ & $345 _{- 32} ^{+ 25}$ & $0.14 _{-0.06}
^{+0.08}$ \nl
174 & $ 58 _{- 51} ^{+ 50}$ & $357 _{- 30} ^{+ 33}$ & $0.14 _{-0.08}
^{+0.08}$ \nl
183 & $ 35 _{- 35} ^{+ 53}$ & $362 _{- 32} ^{+ 31}$ & $0.13 _{-0.08}
^{+0.08}$ \nl
193 & $ 41 _{- 41} ^{+ 41}$ & $373 _{- 28} ^{+ 28}$ & $0.13 _{-0.06}
^{+0.08}$ \nl
203 & $ 41 _{- 41} ^{+ 45}$ & $369 _{- 32} ^{+ 28}$ & $0.12 _{-0.06}
^{+0.09}$ \nl
213 & $ 48 _{- 48} ^{+ 41}$ & $375 _{- 34} ^{+ 26}$ & $0.11 _{-0.06}
^{+0.10}$ \nl
223 & $ 41 _{- 41} ^{+ 50}$ & $380 _{- 31} ^{+ 33}$ & $0.11 _{-0.06}
^{+0.10}$ \nl
233 & $ 62 _{- 51} ^{+ 55}$ & $381 _{- 28} ^{+ 29}$ & $0.10 _{-0.05}
^{+0.09}$ \nl
244 & $ 23 _{- 23} ^{+ 53}$ & $375 _{- 25} ^{+ 31}$ & $0.09 _{-0.06}
^{+0.09}$ \nl
253 & $ 74 _{- 64} ^{+ 61}$ & $366 _{- 26} ^{+ 24}$ & $0.09 _{-0.07}
^{+0.09}$ \nl
264 & $ 45 _{- 45} ^{+ 62}$ & $354 _{- 29} ^{+ 30}$ & $0.07 _{-0.07}
^{+0.10}$ \nl
273 & $ 45 _{- 45} ^{+ 59}$ & $358 _{- 26} ^{+ 33}$ & $0.06 _{-0.06}
^{+0.11}$ \nl
283 & $ 74 _{- 68} ^{+ 51}$ & $359 _{- 28} ^{+ 25}$ & $0.05 _{-0.05}
^{+0.11}$ \nl
294 & $126 _{- 54} ^{+ 49}$ & $362 _{- 27} ^{+ 27}$ & $0.05 _{-0.05}
^{+0.11}$ \nl
304 & $134 _{- 53} ^{+ 50}$ & $356 _{- 24} ^{+ 28}$ & $0.07 _{-0.07}
^{+0.12}$ \nl
314 & $155 _{- 59} ^{+ 52}$ & $344 _{- 30} ^{+ 24}$ & $0.10 _{-0.10}
^{+0.12}$ \nl
324 & $187 _{- 69} ^{+ 53}$ & $354 _{- 30} ^{+ 23}$ & $0.12 _{-0.10}
^{+0.11}$ \nl
333 & $196 _{- 69} ^{+ 57}$ & $355 _{- 27} ^{+ 26}$ & $0.14 _{-0.12}
^{+0.14}$ \nl
344 & $199 _{- 59} ^{+ 56}$ & $362 _{- 33} ^{+ 29}$ & $0.18 _{-0.14}
^{+0.15}$ \nl
353 & $227 _{- 63} ^{+ 53}$ & $368 _{- 32} ^{+ 27}$ & $0.20 _{-0.15}
^{+0.15}$ \nl
363 & $218 _{- 63} ^{+ 63}$ & $364 _{- 38} ^{+ 36}$ & $0.20 _{-0.16}
^{+0.15}$ \nl
373 & $229 _{- 61} ^{+ 61}$ & $373 _{- 41} ^{+ 39}$ & $0.24 _{-0.15}
^{+0.14}$ \nl
383 & $267 _{- 71} ^{+ 59}$ & $381 _{- 45} ^{+ 46}$ & $0.31 _{-0.12}
^{+0.12}$ \nl
388 & $302 _{- 76} ^{+ 62}$ & $380 _{- 50} ^{+ 43}$ & $0.34 _{-0.12}
^{+0.12}$ \nl
\enddata

\tablenotetext{}{
All values were computed for a fixed position angle of
120$^{\circ}$ East of North.\\
The radius is given in arcsec, e.g.~for a distance to M87
of 16.5 Mpc, 1 arcsec would correspond to $\simeq 80$ pc.\\
The velocity dispersion is corrected for rotation and
velocity errors.\\
The listed spin was computed in a 40\arcsec\ bin around each point\\
}
\end{deluxetable}

\clearpage

\onecolumn

\begin{figure}
\psfig{figure=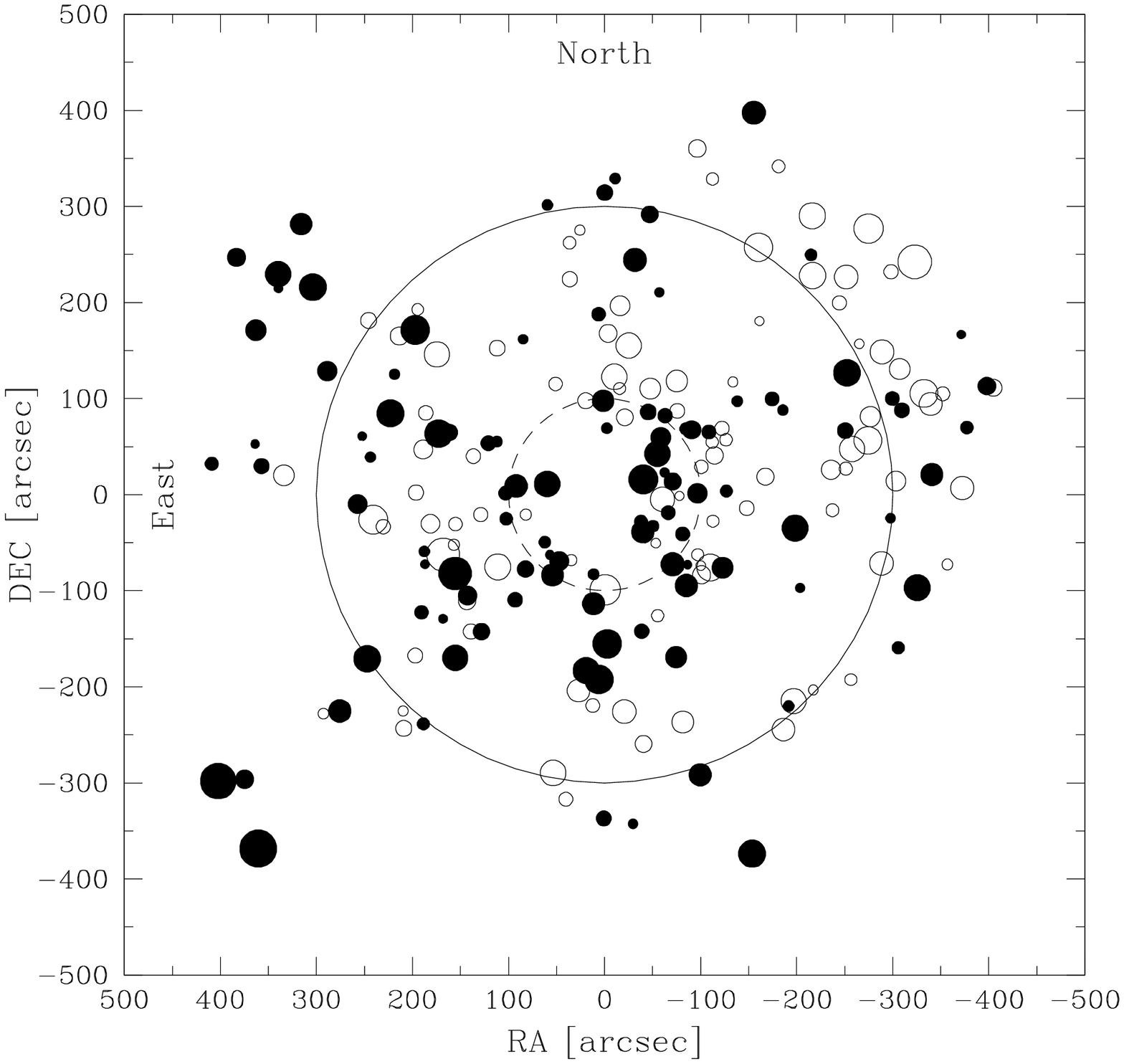,height=16cm,width=16cm
,bbllx=8mm,bblly=57mm,bburx=205mm,bbury=245mm}
\caption{Position of the globular clusters around M87. 
The sizes of the symbols reflect the difference between the globular
cluster velocity and the systemic velocity of the galaxy. Open circles
represent approaching, filled circles represent receding globular clusters; the
dashed and solid lines mark 1 $r_{\rm eff}$  and 3 $r_{\rm eff}$ of the
galaxy respectively.
}
\end{figure}

\clearpage

\begin{figure}
\psfig{figure=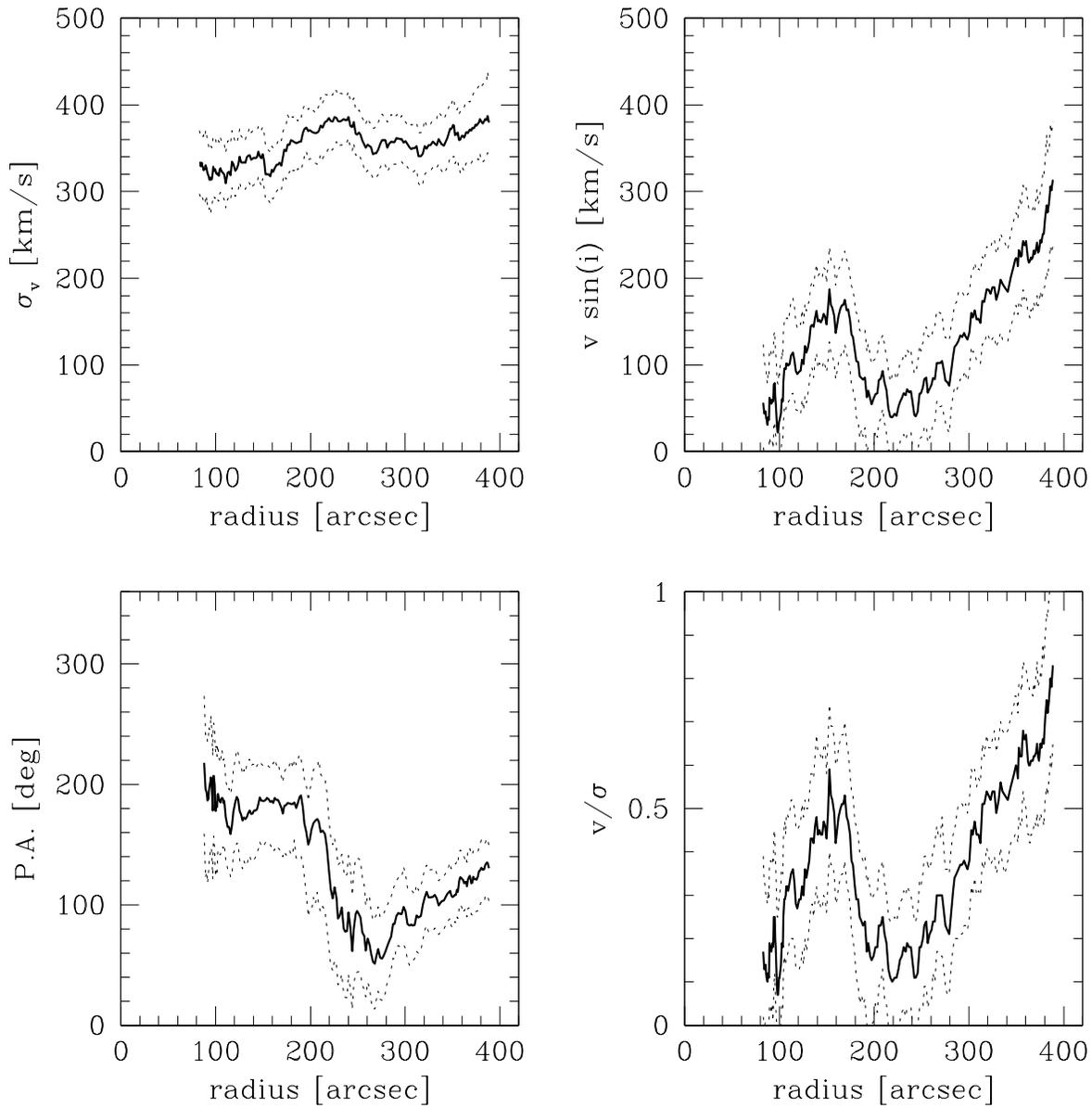,height=16cm,width=16cm
,bbllx=8mm,bblly=57mm,bburx=205mm,bbury=245mm}
\caption{
Projected velocity dispersion, projected rotational velocity, position
angle and $v\cdot sin(i) / \sigma_v$ as functions of radius.
Dotted lines mark the 68\% confidence bands.
}
\end{figure}

\clearpage

\begin{figure}
\psfig{figure=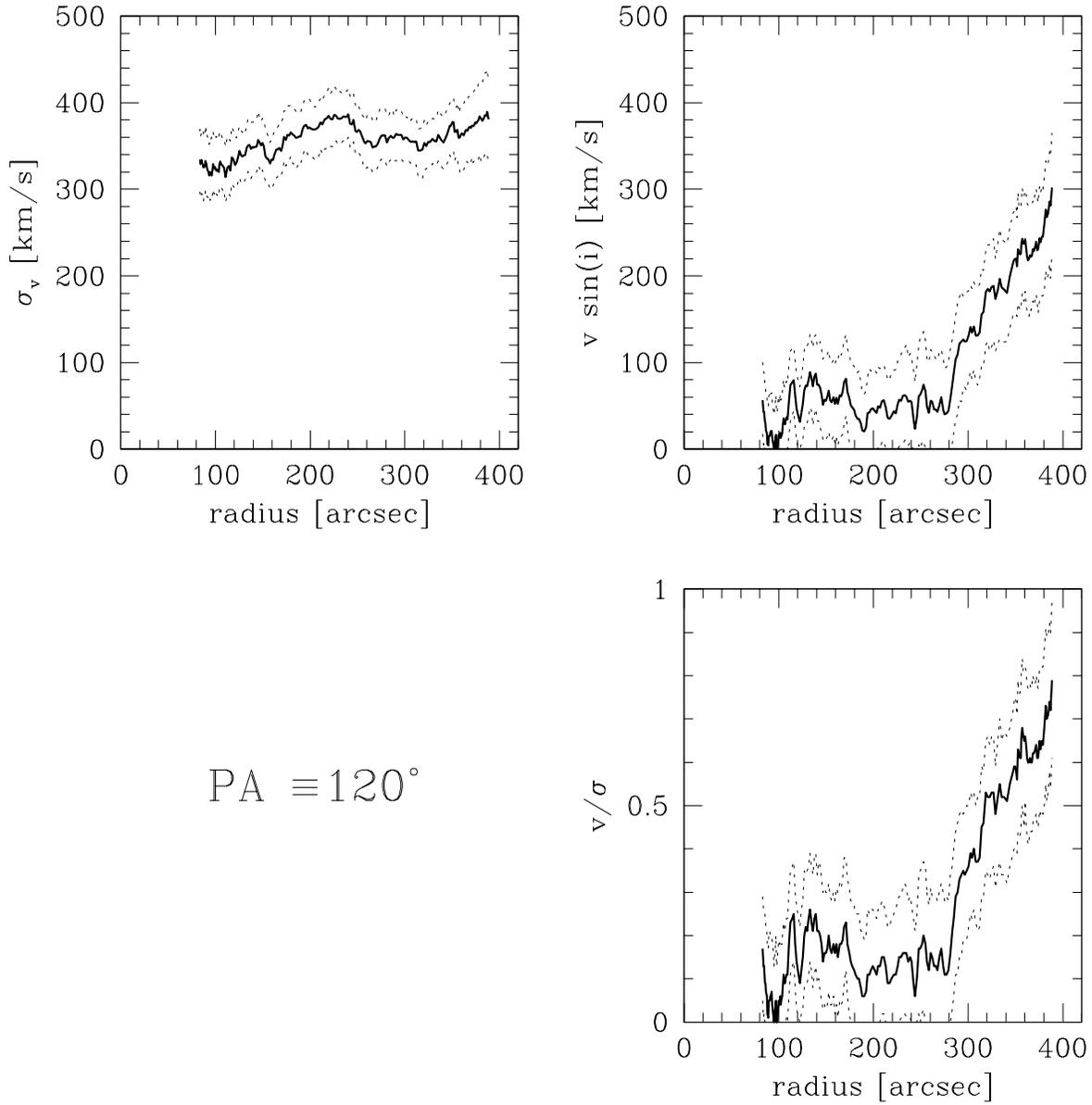,height=16cm,width=16cm
,bbllx=8mm,bblly=57mm,bburx=205mm,bbury=245mm}
\caption{
Projected velocity dispersion, projected rotational velocity, and $v\cdot sin(i) / \sigma_v$
as functions of radius for a fixed position angle of 120$^{\circ}$.
Dotted lines mark the 68\% confidence bands.
}
\end{figure}

\clearpage

\begin{figure}
\psfig{figure=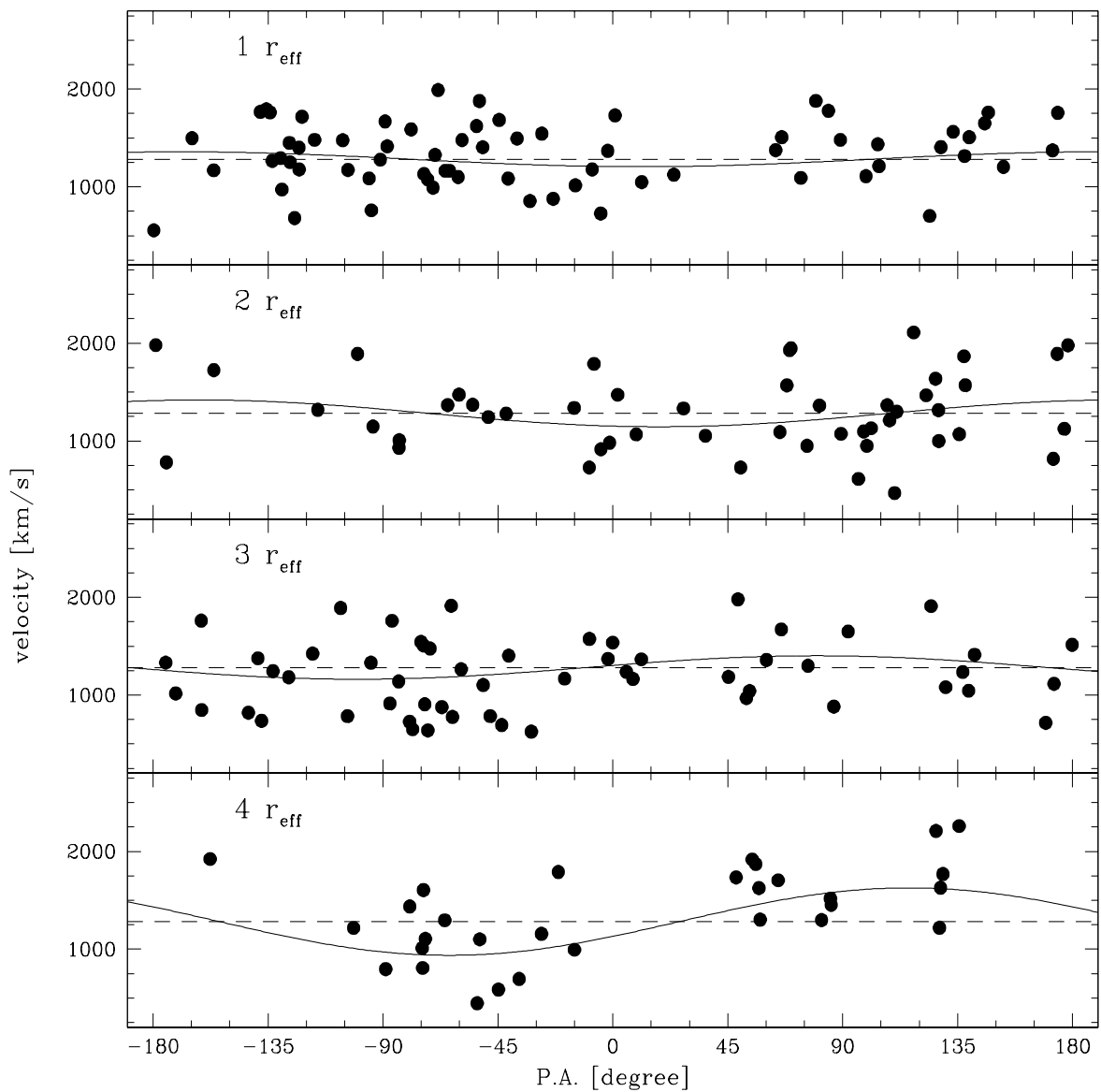,height=16cm,width=16cm
,bbllx=8mm,bblly=57mm,bburx=205mm,bbury=245mm}
\caption{
Velocities of the globular clusters plotted against their position
angles for four radial bins centered on 1, 2, 3, and 4 r$_{\rm
eff}$ (assuming 1 r$_{\rm eff} = 100"$). The solid lines show the best
fitted rotation.
}
\end{figure}

\clearpage

\begin{figure}
\psfig{figure=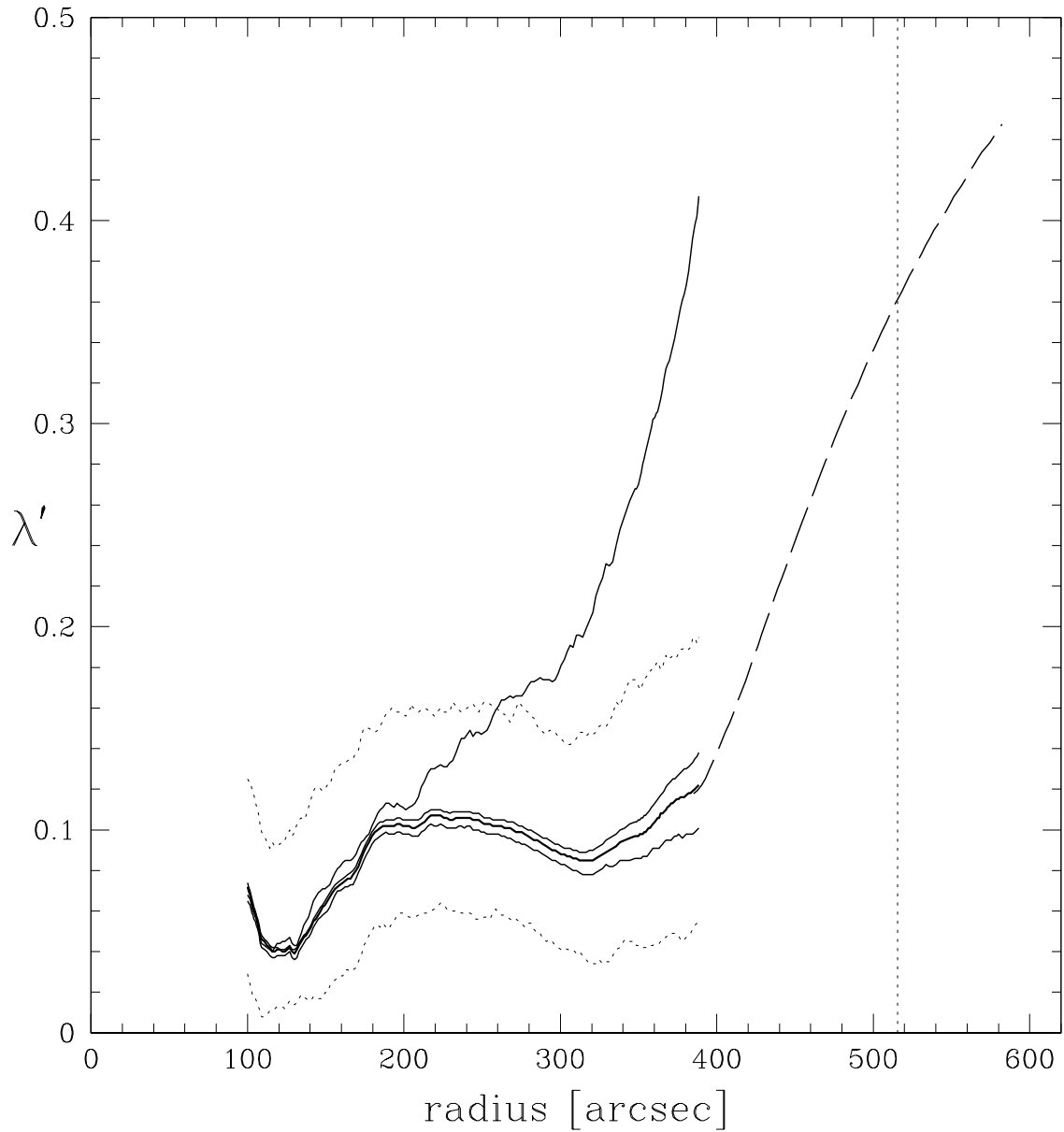,height=16cm,width=16cm
,bbllx=8mm,bblly=57mm,bburx=205mm,bbury=245mm}
\caption{
The dimensionless spin parameter $\lambda^\prime$ (integrated over
radius) plotted against radius. The solid
lines show $\lambda^\prime$ for four different assumptions. In the
extreme (unrealistic) case
where we assumed the rotation velocity to drop to 0 km s$^{-1}$ beyond our 
last data point, $\lambda^\prime$ increase rapidly to very high values beyond
200\arcsec\ .
The three other cases --- rotation velocity stays constant, increases
slowly, or decreases slowly beyond our last data point --- produce very
similar values for $\lambda^\prime$. 
The dotted lines show the 68\% confidence band in the case of
a flat rotation curve beyond our last data point. The last data point of
the sample is indicated as a vertical dotted line.
The long dashed line shows the behavior of $\lambda^\prime$ as
extrapolated from our model.
}
\end{figure}

\clearpage

\begin{figure}
\psfig{figure=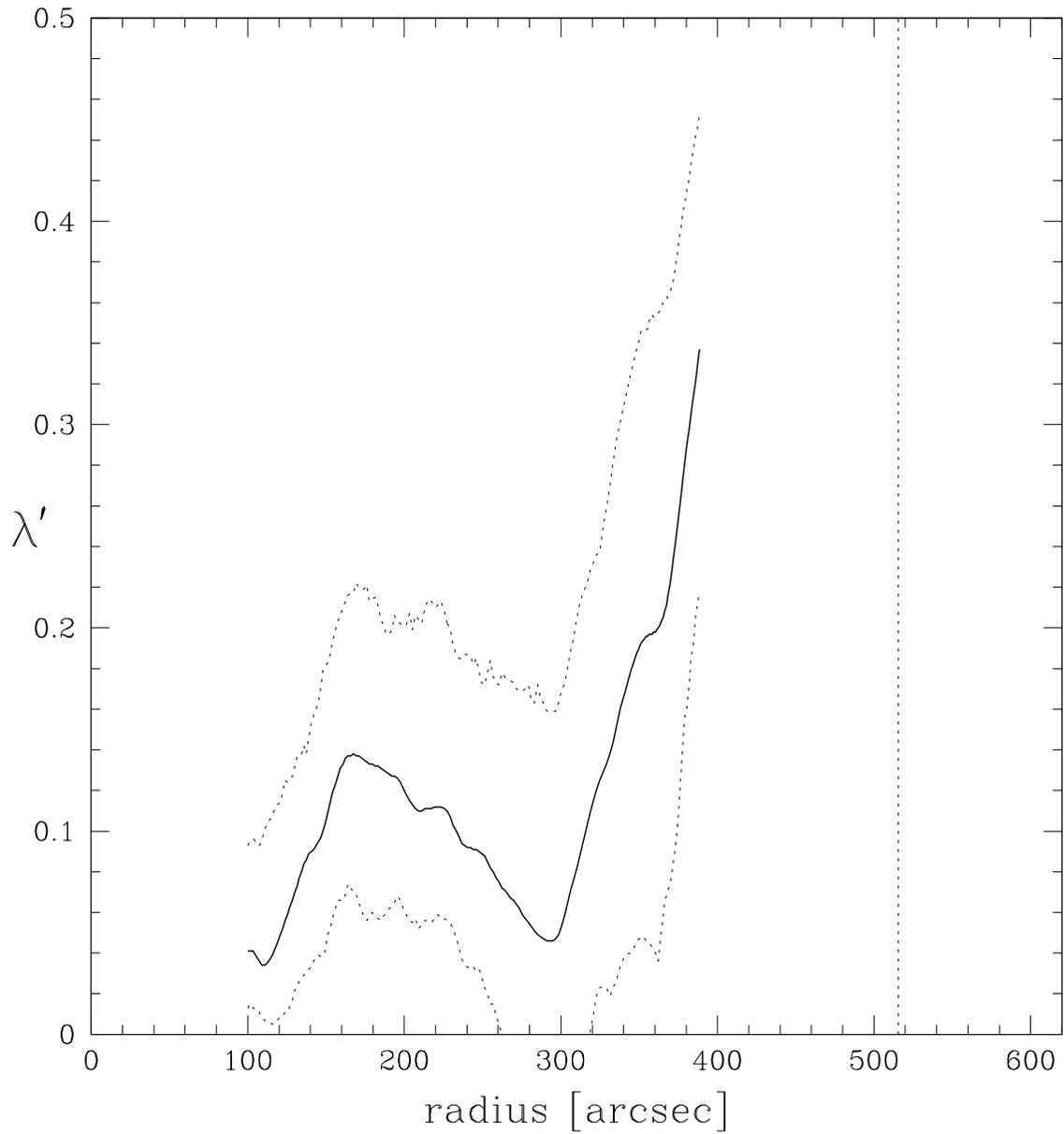,height=16cm,width=16cm
,bbllx=8mm,bblly=57mm,bburx=205mm,bbury=245mm}
\caption{
Dimensionless spin parameter $\lambda^\prime$ computed in 40\arcsec\
bins plotted against radius by assuming a constant rotational 
velocity beyond our last data point.
}
\end{figure}

\clearpage

\begin{figure}
\psfig{figure=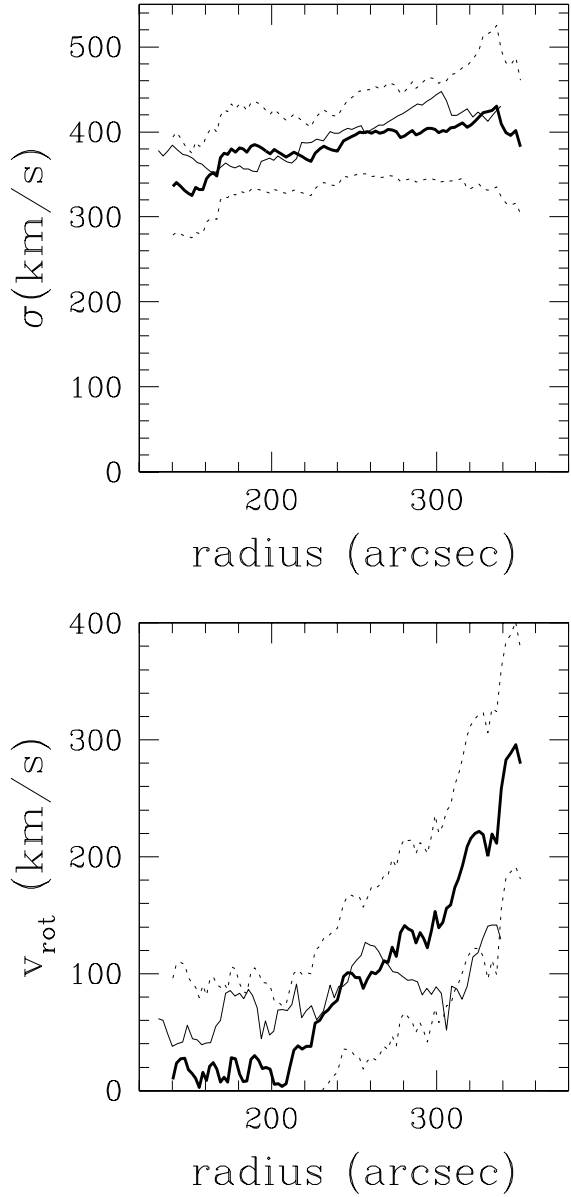,height=16cm,width=16cm
,bbllx=8mm,bblly=57mm,bburx=205mm,bbury=245mm}
\caption{
Projected velocity dispersion (upper panel) and projected rotation velocity 
(lower panel)
plotted against radius for the blue ([Fe/H]$<-0.9$, thick lines) and
red ([Fe/H]$>-0.9$, thin lines) globular clusters. The dotted lines
show the 68\% confidence bands for the blue population; the uncertainties
for the blue and red samples are similar.
}
\end{figure}

\end{document}